\def\papertitle{chowdsp\_wdf: An Advanced C++ Library for Wave Digital Circuit Modelling}
\def\paperauthorA{Jatin Chowdhury}

\documentclass[twoside,a4paper]{article}
\usepackage{dafx_19}
\usepackage{amsmath,amssymb,amsfonts,amsthm}
\usepackage{euscript}
\usepackage[latin1]{inputenc}
\usepackage[T1]{fontenc}
\usepackage{ifpdf}

\usepackage[english]{babel}
\usepackage{caption}
\usepackage{subfig} 
\usepackage{xcolor}
\definecolor{codegreen}{rgb}{0,0.6,0}

\usepackage{listings}
\ninept

\usepackage{times}

\newif\ifpdf
\ifx\pdfoutput\relax
\else
   \ifcase\pdfoutput
      \pdffalse
   \else
      \pdftrue
\fi

\ifpdf 
  \usepackage[pdftex,
    pdftitle={\papertitle},
    pdfauthor={\paperauthorA},
    colorlinks=false, 
    bookmarksnumbered, 
    pdfstartview=XYZ 
  ]{hyperref}
  \usepackage[pdftex]{graphicx}
  \usepackage[figure,table]{hypcap}
\else 
  \usepackage[dvips]{epsfig,graphicx}
  \usepackage[dvips,
    colorlinks=false, 
    bookmarksnumbered, 
    pdfstartview=XYZ 
  ]{hyperref}
  \usepackage[figure,table]{hypcap}
\fi

\usepackage{tikz}
\usetikzlibrary{positioning}
\usepackage[american]{circuitikz}
\usepackage{tkz-euclide}
\usepackage{cleveref}
\usepackage{siunitx}
\usepackage{multirow}

\crefname{lstlisting}{listing}{listings}
\Crefname{lstlisting}{Listing}{Listings}

\DeclareMathAlphabet{\mathpzc}{OT1}{pzc}{m}{it}

\title{\papertitle}

\affiliation{
\paperauthorA \,}
{Chowdhury DSP \\ Denver, CO \\ {\tt \href{mailto:jatin@ccrma.stanford.edu}{jatin@ccrma.stanford.edu}}}

\begin{document}
\ifpdf 
  \DeclareGraphicsExtensions{.png,.jpg,.pdf}
\else  
  \DeclareGraphicsExtensions{.eps}
\fi

\graphicspath{{./Figures/}}
\def\CW{\texttt{chowdsp\_wdf}}
\def\RT{\texttt{RT-WDF}}
\def\WD{\texttt{wdmodels}}
\def\CPP{\texttt{C++}}
\def\FST{\texttt{Faust}}
\def\RTYP{$\mathcal{R}$-Type}
\def\footlink#1{\footnote{\href{#1}{#1}}}

\maketitle
\begin{abstract}
\CW\ is a \CPP\ library for implementing real-time wave digital models of analog circuits. \CW\ differs from existing wave digital modelling libraries by providing a template meta-programming interface for modelling circuits with a fixed topology, and providing support for explicit SIMD acceleration. The motivation and design of the library are described, as well as real-world use-cases, and performance comparisons with other wave digital modelling libraries.
\end{abstract}

\section{Introduction}
Wave Digital Filters (WDFs) are a circuit modelling paradigm originally developed by Alfred Fettweis in the 1970's and '80's \cite{Fettweis}. WDFs have gained popularity in recent years for developing ``virtual analog'' (VA) emulations of audio circuits \cite{KurtThesis}. VA emulations are often implemented as part of software synthesizers or audio effects that are designed to run in real-time, so the run-time performance of these emulations is an important consideration.

\subsection{Wave Digital Filters Background}
While this paper will not attempt to provide a complete introduction to the WDF paradigm, there are several properties of WDFs which should be mentioned. First, WDFs are ``modular,'' meaning that a model of a circuit can be separated into its component parts. This property makes it possible for a wave digital modelling library to implement a single ``Resistor'' element, which can then be re-used in any wave digital circuit model (and so on with most other circuit elements). These element models may also provide interfaces for changing their parameters (e.g. the resistance value of a resistor).
\newline\newline
Wave digital models are comprised of the circuit's element models, along with a set of ``adaptors'' describing how the circuit elements are connected. Common adaptors include ``series'' and ``parallel'' adaptors, as well as polarity inverters. The circuit elements are typically arranged in a tree-like structure, so that each adaptor has a single ``parent'' element and one or more ``child'' elements. When processing signals through the wave digital model, each element requires only information that may be provided by it's child elements. However, when the circuit topology changes, or when a component value is changed (e.g. a potentiometer has been adjusted), it is sometimes necessary for an element to signal the change to its parent element.
\newline\newline
Finally, rather than using traditional circuit quantities such as voltage and current, WDFs operate on wave variables. Most WDF circuit elements accept a single ``incident'' wave and output a single ``reflected'' wave. Adaptors accept $N$ incident waves and output $N$ reflected waves, where $N$ is the total number of child and parent elements connected to the adaptor. The process of computing the output reflected waves from the incoming incident waves for an adaptor may generally be expressed in the form of a scattering matrix, however one-multiply forms of the scattering operation exist for series and parallel adaptors.
\section{Previous Work}
Several libraries for creating real-time wave digital models exist, including \RT, a \CPP\ library published in 2016 \cite{RT_WDF}, and \WD, a \FST\ library published in 2021 \cite{dirk_wdfs}. There are also several libraries not being considered in this paper (e.g. \cite{gustav_anthon_2022_7116075}) since they are written in languages which are not well-suited for real-time audio applications, however, these libraries may be useful for prototyping and for performing non-real-time simulations.
\newline\newline
\RT\ is based on \CPP's concept of run-time polymorphism. Each wave digital element and adaptor is derived from a base class which provides interfaces for all shared WDF functionality. Additionally, \RT\ defines the sample rate as a parameter of the constructor for classes that require knowledge of the system sample rate (capacitors, inductors, etc.). Since the system sample rate may not be known at compile-time, circuit elements are typically constructed dynamically using ``heap'' memory, i.e. with \CPP's \texttt{std::unique\_ptr<>}. For custom adaptors that require more complex scattering matrices, \RT\ interfaces with the Armadillo\footnote{\url{https://arma.sourceforge.net/}} library to perform optimized matrix operations.
\newline\newline
\WD\ is written using the \FST\ programming language. \FST\ is a functional programming language for expressing audio processing algorithms, which may be compiled into \CPP, LLVM Bitcode, \texttt{Rust}, and other programming languages. \FST\ requires the signal flow of the system to be fixed at compilation time, meaning that the topology of the circuit being modelled must be fixed at compile-time as well. \FST\ also does not provide the programmer with mechanisms for interacting with hardware-level instructions, including SIMD intrinsics (helpful for implementing faster matrix operations for adaptors that require more complex scattering matrices), or bit-twiddling (helpful for implementing approximations of various math functions, see e.g. \cite{dangelo_lambertW}).

\section{Library Design}
\CW\ intends to provide the best features of both \RT\ and \WD, with improved performance and flexibility.\footnote{\url{https://github.com/Chowdhury-DSP/chowdsp_wdf}} The library is open-source and is published under the BSD ``3-clause'' license.

\subsection{Run-Time Polymorphism}
For situations where the circuit topology may need to be changed at run-time, \CW\ adopts a similar strategy as \RT, using \CPP's run-time polymorphism, except that components may have their sample rate changed after construction. This property allows circuit elements to be constructed using ``stack'' memory rather than ``heap'' memory, resulting in improved locality of reference \cite{locality}. However, the polymorphic approach has significant performance limitations when compared with \WD's strategy of defining the circuit topology at compile-time (limiting though it may be). Whenever an adaptor in \RT\ needs to access some property of one of its children, it must do so through the polymorphic interface of the child's base class, which is typically implemented as a virtual method table or ``vtable'' \cite{cppreference}. Performing an operation by vtable lookup introduces a small performance overhead and prevents the compiler from performing an inline expansion of the code being executed \cite{inlining}.
\newline\newline
The negative performance effects of implementing a wave digital circuit model using run-time polymorphism can be demonstrated by analyzing the assembly instructions generated by an optimizing compiler. When compiling the simple circuit model shown in listing \ref{lis:VdivDynEx} using the Clang 13.1.6 compiler, with the \texttt{-std=c++20 -O3} compiler flags, the generated assembly for the \texttt{process()} method contains four virtual function \texttt{call} instructions (two for \texttt{S1} to receive the reflected waves from \texttt{R1} and \texttt{R2}, and two to send incident waves to those same elements). These virtual function calls add a small amount of computational overhead and prevent the compiler from performing an inline expansion of the code being run behind the \texttt{call} instructions, which then prevents further optimizations from taking place. For larger circuit models containing more circuit elements, these issues become an increasingly significant performance bottleneck.
\begin{lstlisting}[label={lis:VdivDynEx},captionpos=b,
    caption={\it A voltage divider model using the run-time API.}]
#include <chowdsp_wdf/chowdsp_wdf.h>

namespace wdf = chowdsp::wdf;
struct VoltageDivider {
    wdf::Resistor<float> R1 { 1.0e3f };
    wdf::Resistor<float> R2 { 1.0e3f };
    wdf::WDFSeries<float> S1 { &R1, &R2 };
    wdf::IdealVoltageSource<float> Vin { &S1 };

    inline float process (float x) noexcept {
        Vin.setVoltage (x);
        Vin.incident (S1.reflected());
        S1.incident (Vin.reflected());
        return R2.voltage();
    }
};

float process(float x, VoltageDivider& wdf) {
   return wdf.process (x);
}
\end{lstlisting}
\subsection{Optimizations Via Template Meta-Programming}
\CW\ provides an API for constructing wave digital circuit models using compile-time template meta-programming. While this API does not have the flexibility to construct arbitrary circuit models at run-time, it can provide significant performance improvements for cases where the circuit topology is fixed at compile-time. The fundamental idea behind this design choice is that giving the compiler the maximum possible information about the circuit model will allow the compiler to perform the best possible optimizations for that model. In the compile-time API, resistors, capacitors, and most other ``one-port'' circuit elements are defined almost identically to their run-time counterparts. However, compile-time adaptors are defined with additional template arguments to determine the types of the child elements which are to be connected to the adaptor. Since the adaptor is aware of the types of its child elements at compile-time, all audio processing operations can be performed directly, i.e. without requiring a vtable lookup.
\begin{lstlisting}[label={lis:VdivStaticEx},captionpos=b,
    caption={\it A voltage divider model using the compile-time API.}]
#include <chowdsp_wdf/chowdsp_wdf.h>

namespace wdft = chowdsp::wdft;
struct VoltageDividerT {
    wdft::ResistorT<float> R1 { 1.0e3f };
    wdft::ResistorT<float> R2 { 1.0e3f };
    wdft::WDFSeriesT<float, decltype (R1), decltype (R2)> S1 { R1, R2 };
    wdft::IdealVoltageSourceT<float, decltype (S1)> Vin { S1 };

    inline float process (float x) noexcept {
        Vin.setVoltage (x);
        Vin.incident (S1.reflected());
        S1.incident (Vin.reflected());
        return wdft::voltage<float> (R2);
    }
};

float process(float x, VoltageDividerT& wdf) {
   return wdf.process (x);
}
\end{lstlisting}
Listing \ref{lis:VdivStaticEx} shows the same circuit model as listing \ref{lis:VdivDynEx}, implemented using \CW's compile-time API. When compiling this code with the same compiler and settings, the generated assembly contains zero \texttt{call} instructions, implying that all the necessary interfaces between the adaptor and its child elements have been inlined by the compiler. A further comparison between the performance of \CW's run-time and compile-time API's is provided in Section \ref{sec:perf}.

\subsection{Data Type Abstraction}
\RT\ uses double-precision floating point numbers to store all quantities in the wave digital circuit model. As with all \FST\ code, wave digital models written with \WD\ use single-precision floating point values by default, however, the \FST\ compiler contains optional arguments for using double-precision or quad-precision floating point values instead.
\newline\newline
\CW\ provides a template abstraction for the data type used to store quantities in the circuit. By default, \CW\ supports \CPP's native floating point data types, as well as SIMD wrappers around those floating point data types, via the XSIMD \CPP\ library.\footnote{\url{https://github.com/xtensor-stack/xsimd}} The ability to construct wave digital circuit models using SIMD data types creates several interesting optimization opportunities for implementers of wave digital circuit models.

\subsubsection{Polyphonic Synthesizer}
Consider a polyphonic synthesizer being implemented for a platform that supports Intel SSE or ARM NEON SIMD intrinsics. The synthesizer voices could be implemented using a SIMD data type containing 4 single-precision floating point numbers, providing up to a 4x performance improvement over the same synthesizer using scalar floating point numbers for each voice. Listing \ref{lis:PolyWDF} shows a minimal example of how such a synthesizer voice might be implemented.

\subsubsection{Parallel Circuits}
Constructing wave digital circuit models using SIMD data types may also be useful when emulating devices that contain multiple instances of the same sub-circuit. One example of this phenomenon is the Buchla 259 wavefolder \cite{buchlafolder}, which contains 5 instances of the same ``folder cell'' circuit in parallel with each other. A similar approach could also be used to emulate bucket-brigade device circuits, which contain many capacitor ``bucket'' circuits in series. For example, a bucket-brigade device with 1024 buckets could be split into 4 parallel sub-circuits each containing 256 buckets, arranged such that the output of the first sub-circuit feeds into the input of the second sub-circuit and so on.
\begin{lstlisting}[label={lis:PolyWDF},captionpos=b,
    caption={\it Polyphonic synthesizer voice implemented using a SIMD data type from the XSIMD library.}]
#include <xsimd/xsimd.hpp>
#include <chowdsp_wdf/chowdsp_wdf.h>   

namespace wdft = chowdsp::wdft;
struct SynthVoice {
    using batch = xsimd::batch<float>;
    
    // Resistor controlling the voice's frequency
    wdft::ResistorT<batch> Rpitch { 1.0e3f };
    // Load resistor for the voice circuit
    wdft::ResistorT<batch> Rl { 1.0e3f };
    // Define other components...
    wdft::WDFSeriesT<batch, ...> S5 { ... };
    // Voltage source for the voice circuit
    wdft::IdealVoltageSource<batch,  decltype (S5)> Vsource { S5 };

    void setFrequency (batch pitch) {
        // The resistor value is set as a batch
        // so each voice can have a different pitch
        Rpitch.setResistance (...);
    }

    inline float process() noexcept {
        Vsource.incident (S5.reflected());
        S5.incident (Vsource.reflected());
        const auto Vo = wdft::voltage<batch> (Rl);
        
        // sum of voltages from all voices
        return xsimd::reduce_add(Vo);
    }
};
\end{lstlisting}

\subsection{\RTYP\ Adaptors}
For circuits with topologies that cannot be broken down strictly into series and parallel connections between components, constructing a wave digital model of the circuit may require the use of an \RTYP\ adaptor \cite{Werner2018ModelingCW}. A simple \RTYP\ adaptor may have any number of child elements, and up to one parent element, and uses a scattering matrix to compute the outgoing reflected waves from the incoming incident waves.
\newline\newline
\CW\ contains interfaces for constructing and working with \RTYP\ adaptors, based on the original implementation provided in \cite{sam_thesis}. In order to update the scattering matrix when a ``downstream'' component value changes, the \RTYP\ adaptors in the compile-time API are implemented using a ``template functor'' so that the adaptor can use some custom logic to compute the scattering matrix. Listing \ref{lis:RType} demonstrates a simple circuit model containing an \RTYP\ adaptor with 3 child elements and no parent elements. For performing operations with the scattering matrix \CW\ will optionally use XSIMD to perform SIMD-accelerated matrix operations.
\begin{lstlisting}[label={lis:RType},captionpos=b,
    caption={\it A simple circuit model containing an \RTYP\ adaptor.}]
#include <chowdsp_wdf/chowdsp_wdf.h>   

namespace wdft = chowdsp::wdft;

// Functor to update the scattering matrix
struct ImpedanceCalc {
  template <typename RType>
  static void calcImpedance (RType& R) {
    // get impedances from child elements
    auto [Ra, Rb, Rc] = R.getPortImpedances();
                
    // re-compute the scattering matrix
    R.setSMatrixData({{ ... }});
  }
};

struct RTypeModel {
    // child elements for the RType adaptor
    wdft::ResistorT<float> R1 { ... };
    wdft::ResistorT<float> R2 { ... };
    wdft::ResistorT<float> R3 { ... };
    
    // create the RType adaptor
    using RType = wdft::RootRtypeAdaptor<float,
                                         ImpedanceCalc,
                                         decltype (R1),
                                         decltype (R2),
                                         decltype (R3)>;
    RType adaptor { R1, R2, R3 };
};
\end{lstlisting}

\subsection{Deferring Adaptor Updates}
For circuits with control parameters, a situation may arises when several component values need to be updated at once. In this situation, some computations may become redundant if an adaptor is alerted multiple times of some changes in ``downstream'' circuit elements. This redundancy can affect run-time performance, particularly for adaptors that require relatively large computations in response to changes in downstream elements, as is often the case with \RTYP\ adaptors. To mitigate these performance impacts, \CW\ provides a mechanism for ``deferring'' adaptor updates until after all downstream elements have been updated.

\section{Performance Comparisons} \label{sec:perf}
In order to compare the run-time performance of each modelling library, a variety of circuit models were implemented with each library.\footnote{\url{https://github.com/jatinchowdhury18/wdf-bakeoff}} Chosen circuits include a passive second-order RC lowpass filter (LPF2), a sub-circuit from the Klon Centaur guitar pedal (FF-2), an RC diode clipper, and the tone stack circuit from the '59 Fender Bassman guitar amplifier \cite{Yeh2006DISCRETIZATIONOT}. The LPF2 and FF-2 circuits were chosen as example circuits of varying complexity with wave digital models containing only linear elements and simple adaptors. The diode clipper circuit was chosen as an example circuit containing a single nonlinear element. The Bassman tone stack was chosen as an example circuit requiring an \RTYP\ adaptor. No implementation of the diode clipper circuit was attempted with \RT, since the library does not provide wave domain implementations of diode elements. Each model was also implemented using \CW's compile-time API, as well as \CW's polymorphic run-time API.
\newline\newline
A 1000 second long audio signal at a sample rate of 48 kHz was processed through each model implementation, using a 2018 Mac Mini, with a 3.2 GHz Intel Core i7 CPU. The time needed for each model implementation to process the signal was measured, and is shown in table \ref{tab:perf}. For each circuit model, the \CW\ polymorphic implementation outperforms the \RT\ implementation, with the performance difference widening for the more complicated circuits. Between the two model implementations that are fixed at compile-time, \CW's compile-time API outperforms the \WD\ implementation for all circuit models, excepting the FF-2 circuit.
\newline\newline
The largest difference in performance between the \CW\ compile-time API and the \WD\ library can be seen in the diode clipper circuit. Both libraries use diode implementations based on the Werner et al.'s model \cite{werner_diodes}, which requires evaluation of the Lambert $\mathcal{W}$ function. The large performance difference can likely be ascribed to the fact that \CW\ implements the Lambert $\mathcal{W}$ with a numerical approximation that utilises floating-point bit twiddling \cite{dangelo_lambertW}, while \WD\ uses a Newton-Raphson solver with a fixed number of iterations.
\begin{table}[h]
    \centering
    \resizebox{0.95\columnwidth}{!}{%
        \begin{tabular}{|c|c|c|c|c|}
        \hline
            \multirow{2}{1.4cm}{} & \multirow{2}{2.0cm}{\CW} & \multirow{2}{2.0cm}{\CW\ Poly.} & \multirow{2}{1.5cm}{\WD} & \multirow{2}{1.2cm}{\RT} \\ & & & & \\
        \hline
            \multirow{1}{1.4cm}{LPF2} & \multirow{1}{2.0cm}{\textbf{0.434}} & \multirow{1}{2.0cm}{2.763} & \multirow{1}{1.5cm}{0.797} & \multirow{1}{1.2cm}{2.959} \\
        \hline
            \multirow{1}{1.4cm}{FF-2} & \multirow{1}{2.0cm}{2.763} & \multirow{1}{2.0cm}{6.127} & \multirow{1}{1.5cm}{\textbf{2.126}} & \multirow{1}{1.2cm}{11.73} \\
        \hline
            \multirow{2}{1.4cm}{Diode Clipper} & \multirow{2}{2.0cm}{\textbf{2.041}} & \multirow{2}{2.0cm}{2.091} & \multirow{2}{1.5cm}{7.805} & \multirow{2}{1.2cm}{N/A} \\ & & & & \\
        \hline
            \multirow{2}{1.4cm}{Bassman Tone Stack} & \multirow{2}{2.0cm}{\textbf{0.705}} & \multirow{2}{2.0cm}{2.849} & \multirow{2}{1.5cm}{0.770} & \multirow{2}{1.2cm}{7.978} \\ & & & & \\
        \hline
        \end{tabular}
    }
    \caption{Performance comparison between each wave digital modelling library, showing the amount of time needed for the circuit model to process 1000 seconds of audio at 48 kHz sample rate. The fastest implementation for each circuit is shown in bold.}
    \label{tab:perf}
\end{table}

\section{Examples}
For students and engineers looking to learn how to implement circuit models using \CW, a GitHub repository\footnote{\url{https://github.com/jatinchowdhury18/WaveDigitalFilters}} has been created containing models of several simple circuits (voltage divider, current divider, RC lowpass filter, diode clipper, etc.), as well as several more complex audio circuits, such as the Baxandall tone control circuit \cite{baxandall}, and the Sallen-Key filter \cite{sallen_key}. There are also several open-source examples of \CW\ being used in audio plugins.\footnote{\url{https://github.com/Chowdhury-DSP/BYOD}}\textsuperscript{,}\footnote{\url{https://github.com/Chowdhury-DSP/ChowKick}}

\section{Conclusion}
\CW\ is a \CPP\ library for implementing wave digital circuit models, with separate APIs for construct circuit models with run-time polymorphism, or with compile-time template meta-programming. The library also supports explicit SIMD acceleration, \RTYP\ adaptors, and deferred adaptor updates. Circuit models implemented with \CW\ typically exhibit superior run-time performance when compared with the same circuit model implemented using other wave digital modelling libraries.
\newline\newline
Future work will involve setting up bindings to use the \CW\ library from other programming languages, including Rust, and Javascript (through the web audio API). Additionally, the author will continue to investigate further performance improvements for the library as whole.

\section{Acknowledgments}
The author would like to thank Jingjie Zhang, Kurt Werner, and Dirk Roosenburg for sharing their knowledge about the theory and implementation of wave digital filters, as well as Sam Schachter for his help with implementing \RTYP\ adaptors. Thanks as well to Eyal Amir and Paul Walker for their crucial insights regarding \CPP\ optimization.

\bibliographystyle{IEEEbib}
\bibliography{references}

\end{document}